\documentclass[runningheads]{llncs}
\usepackage{amsmath}
\usepackage{graphicx}
\usepackage{hyperref}
\usepackage{float}
\usepackage{etoolbox}
\usepackage[english]{babel}
\usepackage{subfig}
\usepackage{comment}
\usepackage[colorinlistoftodos]{todonotes}

\newcommand{\fat}[1]{\mathbf{#1}} % Bold for anything else
\newcommand{\bldgr}[1]{\boldsymbol{#1}} % Bold forlower-case Greek letters
\newcommand{\transp}{^T} % Transpose

\newcommand{\argmax}{\operatornamewithlimits{argmax}}
\newcommand{\argmin}{\operatornamewithlimits{argmin}}

\definecolor{noteToSelfColor}{rgb}{0.6,0.6,0.6}
\newcommand{\noteToSelf}[1]{\color{noteToSelfColor}[{#1}]\normalcolor}
\renewcommand{\noteToSelf}[1]{}
\definecolor{todoDoneColor}{rgb}{0.0,0.6,0.0}

\definecolor{unresolvedColor}{rgb}{0.6,0.0,0.0}

\definecolor{questionColor}{rgb}{1.0,1.0,0.0}

\definecolor{answerColor}{rgb}{0.0,1.0,0.0}

\begin{document}

\begin{frontmatter}

%\title{A Contrast-Adaptive Longitudinal Method for Simultaneous Whole-Brain and Lesion Segmentation in Multiple Sclerosis}
\title{A Longitudinal Method for Simultaneous Whole-Brain and Lesion Segmentation in Multiple Sclerosis}

\author{Stefano Cerri\inst{1} \and
Andrew Hoopes\inst{2} \and
Douglas N. Greve\inst{2,3} \and
Mark M\"{u}hlau\inst{4} \and
Koen Van Leemput\inst{1,2}}
%index{Cerri,Stefano}
%index{Hoopes,Andrew}
%index{Greve,Douglas}
%index{M\"{u}hlau,Mark}
%index{Van Leemput,Koen}

\institute{Department of Health Technology, Technical University of Denmark, Denmark
\and
Athinoula A. Martinos Center for Biomedical Imaging, Massachusetts General Hospital, Harvard Medical School, USA
\and
Department of Radiology, Harvard Medical School, United States
\and
Department of Neurology and TUM-Neuroimaging Center, School of Medicine, Technical University of Munich, Germany}

\titlerunning{A Longitudinal Method for Whole-Brain and Lesion Segmentation}
\authorrunning{ }

\pagenumbering{arabic}

\maketitle          

\begin{abstract}
%
%In this paper we propose a longitudinal contrast-adaptive method for simultaneously segmenting anatomical brain structures and white matter lesion in Multiple Sclerosis patients.
%
In this paper we propose a novel method for the segmentation of longitudinal brain MRI scans of patients suffering from Multiple Sclerosis. 
The method builds upon an existing cross-sectional method for simultaneous whole-brain and lesion segmentation, introducing subject-specific latent variables 
to encourage temporal consistency between longitudinal scans. 
%that encourage segmentation across time points to be similar to each other.
%
%The model can adapt to different MRI contrasts and scanners and it does not have any requirements on the number or the timing of longitudinal follow-up scans.
It is very generally applicable, as it does not make any prior assumptions on the scanner, the MRI protocol, or the number and timing of longitudinal follow-up scans.
Preliminary experiments on three 
%different 
longitudinal
datasets indicate that 
%the model  produces more consistent and reliable segmentations compared to the cross-sectional model while detecting better group differences.
the proposed method produces more reliable segmentations and detects disease effects better than the cross-sectional method it is based upon.
\end{abstract}

\end{frontmatter}

\section{Introduction}
\label{sec:Intro}

Multiple Sclerosis (MS) is an inflammatory autoimmune disorder of the central nervous system. It is characterized by the formation of lesions in the white matter, as well as marked brain atrophy primarily in deep gray matter structures~\cite{Barkhof2009,Azevedo2018}. 
The increased availability of longitudinal magnetic resonance imaging (MRI) scans 
%allows 
%enables
opens up the prospect of
tracking lesion evolution and atrophy trajectories over time, 
%thus 
%[ which [should] [may] help assessing ]
%[thereby helping to assess]
enabling a better assessment of
disease progression and treatment efficacy~\cite{Thompson2018}.
%with potential applications in improved assessment of disease progression and treatment efficacy~\cite{Thompson2018}.

% SOTA lesion and limitations
% (Hard to summarize them in 1-2 paragraphs) Carass2017 has a bit of a review
%There have been only few works on longitudinal white matter lesion segmentation 
%Despite its clinical potential, 
Despite high potential clinical impact, 
work on 
computational methods 
for quantifying longitudinal changes in MS 
%[is] 
has remained fairly limited to date
(cf.~\cite{Carass2017} for an overview).
%[\cite{Battaglini2014,Jain2016,Schmidt2019,McKinley2020} ]
%\todo{This needs work. See all these papers for an overview? Why separating Carass from the list? Maybe you only cite an overview paper first, and specific ones in the correct location in the following sentences?}).
%some of them rely on direct image subtraction between two time points
%~\cite{Battaglini2014,Eichinger2017}
%or combined with deformation field
%~\cite{Salem2018}
%\cite{Jain2016}
%instead, use a Expectation-Maximization approach ... Another 
%\cite{Schmidt2019}
%Another approach is to used Convolutional Neural Networks
%\cite{Birenbaum2016,McKinley2020}.
%However, all of these methods have one or more of the following limitations: 1) they segment only white matter lesions 2) they are not contrast adaptive 3) they can compare only two consecutive time points, limiting their usefulness in practice.
%[ Furthermore, ]
%existing methods 
The methods that do exist
suffer from one or more of the following limitations: 
%(1) they segment only white matter lesions (2) they are not contrast adaptive (3) they can compare only two consecutive time points, limiting their usefulness in practice.
They only assess changes in white matter lesions~\cite{Guttmann1999,Gerig2000,Schmidt2019,McKinley2020} 
%\todo{Need some references here -- Guido Gerig had first-author paper in MEDIA ~20y ago; same for David Rey from INRIA Nice; before that Ron Kikinis and Charles Guttmann had a twin paper about longitudinal MS lesion segmentation. I think this are all cited in my TMI 2001 paper}
or in aggregate measures of brain atrophy such as global brain or gray matter volume~\cite{Smith2002,Smeets2016},
%rather than 
but not
in individual brain structures; 
they can only compare between two consecutive time points~\cite{Smith2001,Rey2002,Battaglini2014,Jain2016}
%,Smeets2016}
instead of characterizing entire temporal trajectories;
or 
they are developed and tested 
%on very specific imaging protocols 
in very specific settings
only, with 
%[ substantially ] 
degraded performance when applied to data from 
%different scanners and acquisition protocols, severely limiting their usefulness in the clinic.
different scanners and acquisition protocols~\cite{GarciaLorenzo2013}
%,Valverde2019}
%[ [ severely ] limiting their usefulness in practice. ]
which limits their usefulness in practice.

In order to address these limitations, here
we propose a dedicated model for simultaneously segmenting anatomical brain structures and white matter lesion from longitudinal multi-contrast MRI scans. 
%\todo{Stefano: Maybe here ref to the segmentation figure?}
%
The proposed method
%, based on a previously validated cross-sectional model~\cite{Cerri2020}, 
%is a longitudinal extension of 
builds upon
a contrast-adaptive method for simultaneous whole-brain and lesion segmentation 
that we previously developed and validated~\cite{Cerri2020}.
%is here augmented with a subject-specific atlas~\cite{Iglesias2016} and intensity latent variables that impose respectively shape and intensity constraints for each structure across time points. 
Here we extend this 
%[cross-sectional]
approach to the longitudinal setting by additionally modeling the expected temporal consistency between repeated scans of the same subject, 
using latent variables 
that introduce a statistical dependency between the time points.
%[that impose a statistical dependency between the time points]
%in a manner similar to~\cite{Iglesias2016}.
%
%by additionally modeling the expected temporal consistency between repeated scans of the same subject. 
%
%The longitudinal version introduces the concept of a “subject-specific atlas” within the generative model – a deformation of the population-wide atlas to represent the average subject-specific anatomy across all time points. This subject-specific atlas is an additional latent variable in the model, imposing a statistical dependency between the time points that encourages the corresponding segmentations to be similar to one another. 
%
%In order to adopt this strategy for longitudinal segmentation with SAMSEG, we have further extended the latent variable idea to also enforce temporal consistency in the appearance parts of the model (Gaussian mixture models at each time point should stay close to a latent, subject-specific model). 
%
%
By segmenting both white matter lesions and anatomical brain structures across time, the resulting method enables tracking deep gray matter atrophy trajectories and lesion evolution simultaneously. The model is fully adaptive to different MRI contrasts and scanners, 
%and has no requirement of the number of time points.
%as well as to both 
and does not put 
any 
%requirements 
constraints
%[limitations]
on
the number 
%and 
or
the timing of longitudinal follow-up scans.
To the best of our knowledge, no other method with these capabilities currently exists.

\begin{figure}[t]
    \begin{minipage}{0.72\linewidth}
    \begin{center}
    \includegraphics[width=\textwidth]{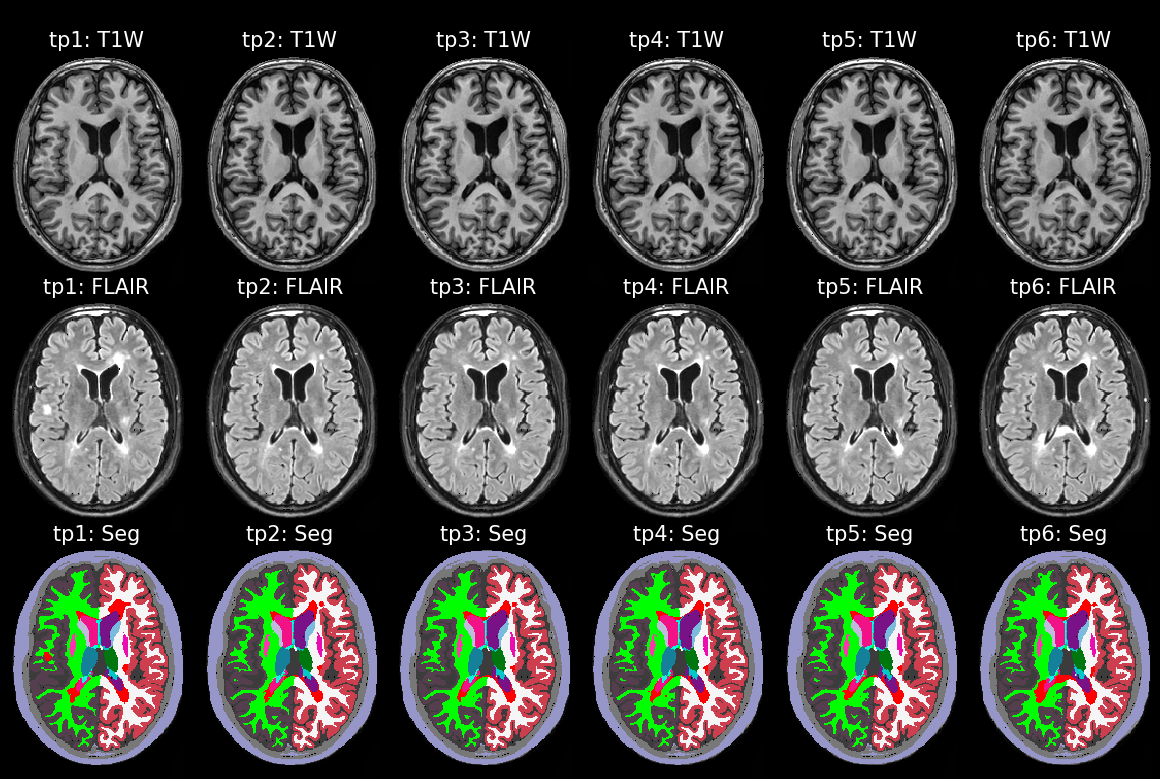}    
    \end{center}
    \end{minipage}
    \begin{minipage}{0.27\linewidth}
        \caption{Example segmentation produced by the proposed method
    %Segmentation
     on a longitudinal scan with T1w and FLAIR contrast. 
    %of the Achieva dataset
    %with T1W and FLAIR as input.
    %\todo{Potentially put caption \emph{beside} figure to save space?}
    }
    \label{fig:segmentationExample}
    \end{minipage}    
\end{figure}

% Comparison/Validation
We assessed the 
%longitudinal
segmentation
performance of the proposed method on three 
%different 
longitudinal
datasets. Preliminary results indicate that it produces more 
%consistent and 
reliable segmentations
%over time 
and detects disease effects better than the cross-sectional method. 
%We validated the model on three different datasets. The model has lower volume differences compared to its cross-sectional version in a test-retest dataset, while it demonstrates better sensitivity in detecting group differences.
%\todo{This needs another pass. Emphasize ``preliminary''}
An example result
produced by the 
longitudinal
method 
is shown in Fig.\ref{fig:segmentationExample}.

%\section{Modeling framework}
%\section{Method}
%\label{sec:ModellingFramework}

% Idea is to refer to the cross sectional method as much as possible while keeping the description of the model understandable. 
% We want to skip all the part relative to the MCMC sampler as is identical to the cross sectional one
% We want to describe here instead the GMM updates
% We need to describe the atlas update for clarity, but we can use Eugenio's paper for ``more details''
% We want (maybe) to also saying somewhere that if we remove the lesion part we get a longitudinal model of the standard SAMSEG

%%Our model is an extension of a previously validated cross-sectional whole-brain and white matter lesion segmentation model~\cite{Cerri2020}. 
%In the following, we first give a brief overview of the existing cross-sectional method
%for simultaneous whole-brain and lesion segmentation~\cite{Cerri2020}.
%%[ which itself is an extension of the [contrast-adaptive] whole-brain segmentation method SAMSEG~\cite{Puonti2016} ]. 
%We then 
%%describe how we extended 
%extend
%this 
%%model 
%approach
%to segment longitudinal scans by
%introducing
%including 
%a hidden subject-specific atlas as well as subject-specific intensity latent variables.
%subject-specific latent variables. 
%[in its model.]
%\todo{In order to save space, this intro could be removed and become a short-sentence intro to section~\ref{subsec:CrossModel}}

\section{Existing cross-sectional method}
\label{subsec:CrossModel}

We first summarize the existing cross-sectional method
for simultaneous whole-brain and lesion segmentation~\cite{Cerri2020} the proposed method builds upon.

Let $\fat{D} = ( \fat{d}_{1}, \dots , \fat{d}_{I} ) $ be the image intensities of a multi contrast MRI scan with $I$ voxels, where the vector $\fat{d}_{i} = ( d_i^1, \dots , d_i^N )\transp $ represents the 
log-transformed
image intensity of voxel $i$ for all the available $N$ contrasts. Moreover, let $\fat{l} = (l_1, \dots , l_I)\transp$ be 
%the corresponding labels, where $l_i \in \{1, \dots,  K\}$ denotes one of the $K$ possible segmentation labels assigned to voxel $i$. 
corresponding segmentation labels, where $l_i \in \{1, \dots,  K\}$ denotes one of the $K$ possible anatomical structures assigned to voxel $i$. 
In order for the model to be capable of segmenting white matter lesions, a binary lesion map $\fat{z} = (z_1, \dots z_I)$ is introduced, where $z_i \in \{0, 1\}$ indicates the presence of lesion in voxel $i$. 
%This binary lesion map is generated by latent code $\fat{h}$ that helps constrain lesion shape. 
We use a  generative model, illustrated in black in Fig.~\ref{fig:model}, 
%\todoDone{Since we're fiddling with notation anyway, I replaced $\cdot_{les}$ with $\cdot_{z}$ everywhere, so same should be done in figure}
to estimate a joint segmentation $\{ \fat{l}, \fat{z} \}$ from MRI data $\fat{D}$.
%\todo{Caption of Fig~\ref{fig:model} needs to explain what black and blue parts mean -- it's the gist of the entire paper}
%\todo{Check the arrows in the figure -- some are wrong}
%The model consists of a segmentation prior $p(\fat{l},\fat{z} | \fat{h}, \fat{x})$ and a likelihood function $p(\fat{D} | \fat{l}, \fat{z}, \bldgr{\theta}, \bldgr{\theta}_{z} )$, where $\fat{h}$, $\fat{x}$, $\bldgr{\theta}$ and $\bldgr{\theta}_{z}$ are model parameters.
%with prior 
%$p(\fat{h})$, $p(\fat{x})$, $p (\bldgr{\theta})$ and $p(\bldgr{\theta}_{z})$.
%$p(\fat{h}) p(\fat{x}) p(\bldgr{\theta}) p(\bldgr{\theta}_{z} | \bldgr{\theta})$.
The model consists of a segmentation prior $p(\fat{l},\fat{z} | \fat{h}, \fat{x})$
with parameters $\fat{h}$ and $\fat{x}$ that encode shape information,
and a likelihood function $p(\fat{D} | \fat{l}, \fat{z}, \bldgr{\theta}, \bldgr{\theta}_{z} )$
%where $\bldgr{\theta}$ and $\bldgr{\theta}_{z}$ are parameters 
with parameters $\bldgr{\theta}$ and $\bldgr{\theta}_{z}$ 
that govern intensity appearance.
%
%Computing an automatic segmentation under this model involves maximizing the posterior distribution of the labels given the data:
%\begin{align}
%\{\hat{\fat{l}}, \hat{\fat{z}} \} = \argmax_{\hat{\fat{l}}, \hat{\fat{z}}} p(\fat{l}, \fat{z} | \fat{D})
%%\propto p (\fat{D} | \fat{l}, \fat{z}) p (\fat{l}, \fat{z})
%.
%\label{eq:posterior}
%\end{align}
%the observed $\fat{D}$ under this model. 
%\todo{The previous two sentences need work: (1) they partly say the same thing; and (2) the first sentence says "The model estimates XYZ using a model", but a model doesn't estimate anything!}
%We now give a brief overview of the segmentation prior and the likelihood function of the model as well as how to obtain automatic segmentation. 
Below we briefly describe the segmentation prior and the likelihood function,
%of the model, 
as well as how 
%we solve this maximization problem.
the model is ``inverted'' to obtain automatic segmentations.
%Eq.~\eqref{eq:posterior}.
%we obtain automatic segmentations [using this model]. 
%\todo{Weird to talk about prior and likelihood without showing what it is. I guess this goes into the previous sentence? How did we solve this in the journal version?}

%\begin{figure}[t]
%    \centering
%    \includegraphics[width=.8\textwidth]{Models/SamsegLongLesionModel.png}
%    \caption{Graphical representation of the proposed model. In black the existing cross-sectional method of~\cite{Cerri2020} for each time point $t$; in blue the proposed additional latent variables for modeling temporal consistency between longitudinal scans with $T$ time points. Shading indicates observed variables. The plate indicates $T$ repetitions of the included variables.
    %\todoDone{$h$ should be $h_t$}
    %\todoDone{Plate and $T$ should also be in blue}
%    }
%    \label{fig:model}
%\end{figure}

\subsubsection{Segmentation prior:} 

%\todo{prior over $\fat{h}$ is missing}

The segmentation prior
is composed of two components
$p(\fat{l} | \fat{x})$ and
$p(\fat{z} | \fat{h}, \fat{x})$
that
%constrains 
encode spatial information regarding 
the neuroanatomical labels 
$\fat{l}$
and 
the lesion map
$\fat{z}$
respectively:
$
  p(\fat{l},\fat{z} | \fat{h}, \fat{x}) 
  =
  p(\fat{l} | \fat{x})
  p(\fat{z} | \fat{h}, \fat{x})
  .
$
%The segmentation prior is composed of two components.
%
The first component is a deformable probabilistic atlas, encoded as a tetrahedral mesh~\cite{VanLeemput2009} with node positions $\fat{x}$ and with a deformation prior distribution defined as:
$$
p(\fat{x}) \propto \exp \left[ -K \sum_d U_{d}(\fat{x}, \fat{x}_{ref}) \right]
    .
$$
%\unresolved{This save a lot of space, but it's really ugly. If possible remove inline eq.}
Here 
$K$ controls the stiffness of the mesh deformations,
$d$ loops over the tetrahedra in the mesh,
%\todo{don't say ``the stiffness parameter'', but rather something that controls the stiffness of the mesh deformations}
and
$U_{d}(\fat{x}, \fat{x}_{ref})$ 
is a cost~\cite{Ashburner2000} 
associated with
%of 
deforming the $d^{th}$ tetrahedron from 
%penalizing the deformation of 
%the $d^{th}$ tetrahedron from 
its shape 
in the atlas's
%population-wide atlas node position 
reference position
$\fat{x}_{ref}$.
Letting
$p( l_i = k | \fat{x} )$ denote the probability of observing label $k$ at voxel $i$ for a given deformation,
assuming conditional independence of the labels between voxels yields
%
%The model assumes conditional independence of the labels between voxels for a given deformation:
%Assuming conditional independence of the labels between voxels for a given deformation, we obtain
%For a given deformation, label probabilities are obtained as
%\begin{align*}
$
  p(\fat{l}| \fat{x}) 
  = \prod_{i=1}^{I} 
    p( l_i | \fat{x} )
  .
  %,
  %\quad
  %p( l_i=k | \fat{x} ) = \sum_j^J \alpha_j^k \psi_j^i(\fat{x})
$  
%\end{align*}
%where 
%$p( l_i | \fat{x} )$ evaluates the label probabilities at voxel $i$ given the node positions $\fat{x}$. 
%$p( l_i = k | \fat{x} )$ is the probability of observing label $k$ at voxel $i$.
%using linear interpolation. 
%where $\alpha_j^k$ are probabilities of label $k$ defined at the $J$ vertices of the mesh, and $\psi_j^i(\fat{x})$ denotes a spatially compact, piecewise-linear interpolation basis function attached to the $j^{th}$ vertex and evaluated at the $i^{th}$ voxel~\cite{VanLeemput2009}.
%\todo{I don't think we need to explain interpolation model here with equations.}

The second component of the segmentation prior is a 
%factorized 
model of the form: 
%\todo{If in desparate need of space, inline this one as not mentioned again anywhere else in the paper}
$
p(\fat{z} | \fat{h}, \fat{x}) 
= \prod_{i=1}^I p( z_i | \fat{h}, \fat{x})
, \textbf{ }%\quad
p ( \fat{h} ) = \mathcal{N} ( \fat{h} | \fat{0}, \fat{I})
,
$
where
$
p( z_i = 1 | | \fat{h}, \fat{x} )
$
is the probability that voxel $i$ is part of a lesion.
This model
%[ combines ]
takes into account 
both
a voxel's spatial location within its neuroanatomical context (through $\fat{x}$),
as well as  
lesion shape constraints
%[ implemented ]
%[ provided by ]
through
a variational autoencoder (VAE)~\cite{Kingma2013} 
%in which
that ``decodes''
a low-dimensional
latent
code 
%variable
$\fat{h}$ 
%is ``decoded'' 
using a convolutional neural network.

\subsubsection{Likelihood:}

For the likelihood, which links segmentations $\{\fat{l}, \fat{z}\}$ to intensities $\fat{D}$, we use a multivariate Gaussian intensity model for each structure,
and model the MRI bias field artifact as a linear combination of spatially smooth basis functions that is added to the local voxel intensities \cite{Wells1996,VanLeemput1999}.
Letting 
%$\bldgr{\theta}_{z}$ 
$\bldgr{\theta}_{z} = \{\bldgr{\mu}_{z}, \bldgr{\Sigma}_{z}\}$ 
%collect 
denote
the mean and variance of lesion intensities, 
and $\bldgr{\theta}$ 
the collection of
%the 
bias field parameters and 
%the
%Gaussian 
intensity
means
and variances 
$\{ \bldgr{\mu}_k, \bldgr{\Sigma}_k \}$
of all 
%other 
$K$
anatomical
structures, the likelihood is defined as
$
  p(\fat{D} | \fat{l}, \fat{z}, \bldgr{\theta}, \bldgr{\theta}_{z} ) 
  = \prod_{i=1}^{I} p(\fat{d}_{i} | l_{i}, z_i, \bldgr{\theta}, \bldgr{\theta}_{z} ),
$
where
\begin{align*}
  p(\fat{d}_{i} | l_i=k, z_i, \bldgr{\theta}, \bldgr{\theta}_{z} ) 
  = 
  \begin{cases}
    \mathcal{N}( \fat{d}_{i} | \bldgr{\mu}_{z} + \fat{C}\transp\bldgr{\phi}_{i}, \bldgr{\Sigma}_{z} ) & \text{if } z_i=1, \\
    \mathcal{N}( \fat{d}_{i} | \bldgr{\mu}_k + \fat{C}\transp\bldgr{\phi}_{i}, \bldgr{\Sigma}_k ) & \text{otherwise}.
  \end{cases}
\end{align*}
Here $\bldgr{\phi}_{i}$ evaluates the bias field basis functions at the $i^{th}$ voxel, and $\fat{C} = (\fat{c}_1 , \dots , \fat{c}_N)$, where $\fat{c}_n$ 
%\todo{(1) vectors should be boldface; and (2) the dimensionality of $\fat{C}$ doesn't match the one of $\bldgr{\phi}_i$}
denotes the parameters of the bias field model for the $n^{th}$ contrast.
The model is completed by a flat prior on $\bldgr{\theta}$, and a weak 
conditional
prior $p(\bldgr{\theta}_{z} | \bldgr{\theta})$ that ensures that the method can be robustly applied to scans with no or very small lesion loads~\cite{Cerri2020}.

%We use a flat prior on $\bldgr{\theta}$, while we constraint the possible values of $\bldgr{\theta}_{z}$ by making them conditional on the white matter Gaussian parameters using a conjugate prior. 
%A conjugate prior
%\todo{(1) flat prior on $\bldgr{\theta}$ missing; and (2) I don't think we need to detail here the exact form of the conditional prior on $\bldgr{\theta}_{z}$, just the gist: why we use it (robustness against low/no lesion load), and that it depends on WM parameters (in words)}
%of the form of a Normal Inverse Wishart distribution is then used to make $\bldgr{\theta}_{z}$ conditional on the other intensity parameters:
%\begin{align}
%   p( \bldgr{\theta}_{z} | \bldgr{\theta} )
%   = 
%   \mathcal{N}( \bldgr{\mu}_{z} | \bldgr{\mu}_{WM}, \nu^{-1}\bldgr{\Sigma}_{z} ) \mathrm{IW}( \bldgr{\Sigma}_{z} | \kappa \nu \bldgr{\Sigma}_{WM}, \nu-N-2 )
%   ,
%   \label{eq:NIWLesion}
%\end{align}
%where ``WM'' denotes the white matter Gaussian and $\kappa > 1$ and $\nu \ge 0$ are hyperparameters in the model.

\subsubsection{Segmentation:}
%\todo{This section needs a thorough revisit. Correct structure would be (1) Aim is to use $p( \fat{l}, \fat{z} | \fat{D})$ -- perhaps this already needed to be explained before detailing the prior and likelihood; (2) Approximation by pluggin point estimates of $\bldgr{\theta}$ and $\fat{x}$; (3) (maybe skip this one) obtaining point estimate involves simplifying the model by clamping decoder output / removing VAE from model; and (4) how exactly (1) is solved doesn't need to be described here, just say involves MCMC/Gibbs sampling}
%
%\todo{Current technical errors: (1) $\hat{\bldgr{\theta}}_{z}$ is estimated but never used since to condition the segmentation posterior, as it's sampled over; (2) clamping ``weights'' of $f_i(\fat{h})$ is meaningless because the reader doesn't know what those weights are (never mentioned!)}
%

Given an MRI scan $\fat{D}$, segmentation proceeds by 
%inferring 
approximating
the segmentation posterior 
using point estimates of the parameters 
$\fat{x}$ and $\bldgr{\theta}$:
\begin{equation}
  p( \fat{l}, \fat{z} | \fat{D} ) 
  \simeq
  p ( \fat{l}, \fat{z} | \fat{D}, \hat{\bldgr{\theta}}, \hat{\fat{x}} )
  \label{eq:segmentationPosteriorApproximation}
  ,
\end{equation}
and Markov chain Monte Carlo sampling to marginalize over the 
remaining, 
lesion-specific parameters $\bldgr{\theta}_{z}$ and $\fat{h}$.
%In order to find
For the purpose of finding
the point estimates 
%of $\fat{x}$ and $\bldgr{\theta}$,
$\fat{\hat{x}}$ and $\bldgr{\hat{\theta}}$,
a simplified model is fitted to the data:
\begin{equation}
  \bldgr{\hat{\Omega}}
  = 
  \argmax_{ \bldgr{\Omega} } 
  p ( \bldgr{\Omega} | \fat{D})
  \quad
  \mathrm{with}
  \quad
  \bldgr{\Omega} = \{ \fat{x}, \bldgr{\theta}, \bldgr{\theta}_{z} \}
  \label{eq:crossSectionalParameterOptimization}
  ,
\end{equation}
where the lesion-shape encoding VAE and its parameters $\fat{h}$ are 
temporarily
removed to simplify the optimization process. 
More details can be found in~\cite{Cerri2020}.

\section{Longitudinal extension}

%In order to make the cross-sectional model capable of handling longitudinal scans, we augment its generative model by introducing subject specific latent variable in the model. 
%
%\todo{``Big picture'' things that are missing: optimization of $\fat{x}_0$ is similar in complexity/method as $\fat{x}$ due to symmetry in deformation cost (cf. Eugenio)}
%

\noindent
%We now 
In the longitudinal setting we
are given $T$ 
%longitudinal 
scans with image intensities 
%$\fat{D} = ( \fat{D}_1, \dots, \fat{D}_T ) $, 
$\{ \fat{D}_t \}_{t=1}^T$,
and we wish to compute for each time point $t$ the corresponding segmentation $\{\fat{l}_t, \fat{z}_t \}$.
In contrast to the cross-sectional setting where each image 
%is 
%[ needs to be analyzed ]
is 
%[ viewed ] 
treated
independently,
here we can exploit
the fact that all images belong to the same subject 
%in the longitudinal setting 
%can be exploited 
to produce more consistent (and potentially more accurate) segmentations.
%
%Eugenio: ``Longitudinal segmentation algorithms exploit the prior knowledge that a set of images belongs to the same subject, in order to produce more accurate and consistent segmentations than when the images are processed independently.''
%
%
%by additionally modeling the expected temporal consistency between repeated scans of the same subject. 
%
Towards this end, we 
%[ propose to ] 
%include additional 
introduce subject-specific
latent variables
$\fat{x}_0$ and $\bldgr{\theta}_0$ 
in the segmentation prior and likelihood, respectively, 
imposing a statistical dependency between the time points that encourages the %corresponding 
segmentations to be similar to one another.
%
%The cross-sectional model described before treats each time point independently; In the proposed method, instead, we introduced a subject-specific atlas, with node positions $\fat{x}_0$, that enforces spatial constraints across time points and intensity latent variables $\bldgr{\theta}_0$ that enforce intensity constraints across time points. 
%
%
The augmented 
generative
model is depicted in Fig.~\ref{fig:model}, where 
the parameters $\fat{x}_t, \fat{h}_t, \bldgr{\theta}_t$ and $\bldgr{\theta}_{t,z}$ denote the model parameters of time point $t$, and
the blue parts indicate the additional components compared to the cross-sectional model.
%
%In the following, we describe the resulting  longitudinal segmentation prior and likelihood, and how we use those to obtain automatic segmentations. \todo{Last sentence perhaps a bit superfluous?}

%\subsubsection{Subject-specific atlas:}
\subsubsection{%[ Longitudinal ] 
Segmentation prior:}

%Instead of using an independent  atlas deformation 
%%prior Eq.~\eqref{eq:XXX} 
%for each time point,
In order to obtain temporal consistency in the segmentation prior,
%constraint shape across time points with
%spatial constraints across time points
we 
%make use of 
use
the concept of
a ``subject-specific atlas''~\cite{Iglesias2016}: a deformation of the
%normal 
cross-sectional
%population-wide 
atlas 
to represent the average subject-specific anatomy across all time points.
In particular, 
$$
  p( \{ \fat{x}_t \}_{t=1}^T | \fat{x}_0 )
  =
  \prod_{t=1}^T  p( \fat{x}_t | \fat{x}_0 )
  ,
  \quad
  p( \fat{x}_{t} | \fat{x}_{0} ) \propto \exp \left[ - K_{1} \sum_d U_{d}(\fat{x}_{t}, \fat{x}_{0}) \right]
  ,
$$
where $\fat{x}_0$ 
%is a latent variable 
are latent atlas node positions 
encoding subject-specific brain shape,
%that 
%[ which is ]
%itself 
%themselves governed by
with prior
$
p( \fat{x}_{0} ) \propto \exp \left[ - K_{0} \sum_d U_{d}(\fat{x}_{0}, \fat{x}_{ref}) \right]
.
$
%\todoDone{Stefano: This save a lot of space, and in my opinion is ok since we already defined the same stuff in the cross-sectional model}
Here 
the mesh stiffnesses
$K_0$ and $K_1$ are hyperparameters of the model;
%Note that 
%for 
by choosing
%$K_0 \rightarrow \infty$ and $K_1 \rightarrow K$ 
$K_0 = \infty$ and $K_1 = K$ 
%the cross-sectional segmentation prior is retained for each of the $T$ time points.
the model devolves into the cross-sectional segmentation prior for each time point separately.

%the model devolves into $T$ independent deformation priors 
%reduces 

%a ``subject-specific atlas'' within the generative model -- a deformation of the population-wide atlas to represent the average subject-specific anatomy across all time points.

%We constraint shape across time points with a subject-specific atlas~\cite{Iglesias2016}.
% From Eugenio's paper, it needs to be rewritten
%We now have a population-wide atlas that is first deformed to a subject-specific atlas
%\begin{align*}
%    p( \fat{x}_{0} ) \propto \exp \left[ - K_{0} \sum_d U_{d}(\fat{x}_{0}, \fat{x}_{ref}) \right]
%    .
%\end{align*}
%The mesh in position $\fat{x}_{0}$ is further deformed $T$ times to positions ${x_{1}, \cdots , x_{T}}$ (corresponding to the $T$ time points), but this time using $x_{0}$ as reference position:
%\begin{align*}
%    p( \fat{x}_{t} | \fat{x}_{0} ) \propto \exp \left[ - K_{1} \sum_d U_{d}(\fat{x}_{t}, \fat{x}_{0}) \right]
%    ,
%\end{align*}
%for $t = 1 \dots T$.

%Note that no assumptions are made a priori on the values of each deformation.

%\subsubsection{Subject-specific intensity latent variables.}
\subsubsection{%[ Longitudinal ] 
Likelihood:}

%We introduce subject-specific intensity latent variables that encourage Gaussian models across time points to be similar across time, without defining a priori their values. 
In a similar vein, we also introduce subject-specific latent variables to encourage
temporal consistency in the
Gaussian intensity models.
%to be similar across time. 
%
For each anatomical structure,
%$k$
%$k = 1, \ldots, K$,
%and for each time point 
we condition
the Gaussian parameters
$\{ \bldgr{\mu}_{t,k}, \bldgr{\Sigma}_{t,k} \}$
%[ of each time point ]
on latent variables 
%$\bldgr{\theta}_{0,k} = \{ \bldgr{\mu}_{0,k}, \bldgr{\Sigma}_{0,k} \}$
$\{ \bldgr{\mu}_{0,k}, \bldgr{\Sigma}_{0,k} \}$
%with form of a 
using a
normal-inverse-Wishart (NIW) distribution:
$
  p( \{ \bldgr{\theta}_{t} \}_{t=1}^T | \bldgr{\theta}_{0} )
  =
  \prod_{t=1}^T  p( \bldgr{\theta}_{t} | \bldgr{\theta}_{0} )
$
with
\begin{align*}
   %p( 
   %\bldgr{\theta}_{t,k}
   %%\bldgr{\mu}_{t,k}, \bldgr{\Sigma}_{t,k} 
   %| \bldgr{\theta}_{0,k} )
   %= 
   p( 
   %\{
   \bldgr{\theta}_{t} 
   %\}_{t=1}^T
   | \bldgr{\theta}_{0} )
   \propto
   %\prod_{t=1}^T 
   \prod_{k=1}^K
   \mathcal{N}( \bldgr{\mu}_{t,k} | \bldgr{\mu}_{0,k}, P_{0,k} \bldgr{\Sigma}_{0,k} ) \mathrm{IW}( \bldgr{\Sigma}_{t,k} | P_{0,k} \bldgr{\Sigma}_{0,k}, P_{0,k} - N - 2 )
   .
   \label{eq:NIW}
\end{align*}
Here 
$\bldgr{\theta}_0 = \{ \bldgr{\mu}_{0,k}, \bldgr{\Sigma}_{0,k} \}_{k=1}^K$
with 
%a flat 
prior $p(\bldgr{\theta}_0) \propto 1$,
and
$P_{0,k} \ge 0 $ is a hyperparameter that governs the strength of the regularization across time for label $k$. Note that choosing $P_{0,k} = 0, \forall k$ yields the cross-sectional likelihood for each time point independently.
%
%[ Note that the Gaussian parameters for lesions do not have any subject-specific latent variables in our model, and are conditioned on the other likelihood parameters as in the cross-sectional model.]
%\todo{not sure we need this; especially because (1) it's already clear from the figure and the equations, and (2) we don't explain why we treat at differently. We also don't say anything special about bias field parameters.}
%and its parameters for each time point are conditioned as in the cross-sectional model.
%in Eq~\eqref{eq:NIWLesion}.

%\subsubsection{Segmentation:}
\subsubsection{%[Longitudinal] 
Segmentation:}
We follow the same overall segmentation strategy as in the cross-sectional setting: 
we first compute point estimates
%$\bldgr{\hat{\theta}}_t$ and $\fat{\hat{x}}_t$ for each time point $t$ 
$\{\bldgr{\hat{\theta}}_t, \fat{\hat{x}}_t \}_{t=1}^T$ 
using a simplified model in which 
the lesion shape codes $\{ \fat{h}_t \}_{t=1}^T$ are removed,
and subsequently obtain segmentations as described in the cross-sectional setting, i.e., by using \eqref{eq:segmentationPosteriorApproximation} for each time point separately. 
As in the cross-sectional case, 
we obtain the required point estimates by fitting the longitudinal model to the data:
\begin{equation}
  \{ \bldgr{\hat{\Omega}}_1, \ldots, \bldgr{\hat{\Omega}}_T, \bldgr{\hat{\theta}}_0, \fat{\hat{x}}_0 \} 
  = 
  \argmax_{ \{ \bldgr{\Omega}_1, \ldots, \bldgr{\Omega}_T, \bldgr{\theta}_0, \fat{x}_0 \} } 
  p ( \bldgr{\Omega}_1, \ldots, \bldgr{\Omega}_T, \bldgr{\theta}_0, \fat{x}_0 | \fat{D}_1, \ldots, \fat{D}_T )
  \label{eq:longitudinalParameterOptimization}
  ,
\end{equation}
where $\bldgr{\Omega}_t = \{ \fat{x}_t, \bldgr{\theta}_t, \bldgr{\theta}_{t,z} \}$.
For optimizing \eqref{eq:longitudinalParameterOptimization} we use coordinate ascent, updating one variable at a time in an iterative fashion. 
Because 
%[ the longitudinal segmentation prior ]
$
p( \fat{x}_t | \fat{x}_0 )
$
is of the same form as the cross-sectional deformation prior,
and the NIW distribution used in
$
p( \bldgr{\theta}_t | \bldgr{\theta}_0 )
$
is the conjugate prior 
for the mean and variance of a Gaussian distribution,
%updating 
estimating
$\bldgr{\Omega}_t$ 
from $\fat{D}_t$
for given values of $\bldgr{\theta}_0$ and $\fat{x}_0$
%is no different than optimizing 
simply involves performing an optimization of the form of
\eqref{eq:crossSectionalParameterOptimization} 
%[ in the cross-sectional setting ]
for each time point $t$ separately.
%and we therefore use the same optimization algorithm.
Conversely, 
for given values 
%$\bldgr{\Omega}_1, \ldots, \bldgr{\Omega}_T$ 
$\{ \bldgr{\Omega}_t \}_{t=1}^T$ 
%updates for $\bldgr{\mu}_{0,k}$ and $\bldgr{\Sigma}_{0,k}$ are 
the update for $\bldgr{\theta}_0$ is
given in closed form:
\begin{align*}
  \bldgr{\mu}_{0,k} 
  \gets  
  \left(\sum_{t=1}^T \bldgr{\Sigma}_{t,k}^{-1} \right)^{-1} \sum_{t=1}^T \bldgr{\Sigma}_{t,k}^{-1} \bldgr{\mu}_{t,k}
  ,
  \quad
    \bldgr{\Sigma}^{-1}_{0,k} 
  \gets 
  \left(\frac{1}{T} \sum_{t=1}^T \bldgr{\Sigma}_{t,k}^{-1} \right)
  \frac{P_{0,k}}{P_{0,k} - N - 2}
  ,
\end{align*}
%i.e., a weighted average of the various $\bldgr{\mu}_{t,k}$'s, each with weight 
%$\left[\sum_{t=1}^T \bldgr{\Sigma}_{t,k}^{-1} \right]^{-1} \sum_{t=1}^T \bldgr{\Sigma}_{t,k}^{-1}$.
%Optimizing for $\bldgr{\Sigma}_{0,k}$ yields to:
%\begin{align*}
%  \bldgr{\Sigma}^{-1}_{0,k} 
%  \quad \gets \quad
%  \left(\frac{1}{T} \sum_{t=1}^T \bldgr{\Sigma}_{t,k}^{-1} \right)
%  \frac{P_{0,k}}{P_{0,k} - N - 2}
%\end{align*}
%thus we are essentially ``averaging'' the precisions.
%As shown already in~\cite{Iglesias2016}, optimizing $\fat{x}_0$ is done by optimizing
whereas updating $\fat{x}_0$ involves 
the optimization
(cf.~\cite{Iglesias2016})
$$
\argmin_{\fat{x}_0} \sum_{d} 
    \left[ 
     K_{0} U_{d}(\fat{x}_{0}, \fat{x}_{ref}) +
     %K_{1} U_{d}(\fat{x}_{t}, \fat{x}_{0})
     K_{1} 
     \sum_{t=1}^T
     U_{d}(\fat{x}_{0}, \fat{x}_{t})
    \right]
    ,
$$
% Eugenio's paper
%The equation can be seen as a weighted ``average'' of the mesh positions of the time points and that of the population-wide atlas $\fat{x}_{ref}$. 
%The atlas essentially plays the role of an additional time point, thought with a different weight ($K_0$, rather than $K_1$).
%We solve this problem numerically with a conjugate gradient algorithm.
which we solve numerically using a limited-memory BFGS algorithm.

\subsubsection{Implementation:}
\label{sec:TrainingAndTuning}

%In order to ensure that no spurious processing biases~\cite{Reuter2011}.

In order to avoid 
longitudinal processing biases %~\cite{Reuter2011}
%spurious biases in the longitudinal processing~\cite{Reuter2011},
%resulting from applying certain processing steps to some time points only (e.g., resampling follow-up scans to a baseline scan),
resulting from not treated all time points in exactly the same way,
%\cite{Reuter2011},
we 
%make use of 
first compute
an unbiased
within-subject template 
%created with 
using
an inverse consistent registration method~\cite{Reuter2012}. This template is a robust representation of the average subject anatomy 
over
%[across] 
time,
and 
%serves
we use it
as an unbiased reference to register all time 
%point data 
points
to in a preprocessing step.
We also use it to 
%initialize 
start
the proposed iterative 
%[ coordinate ascent ]
algorithm optimizing \eqref{eq:longitudinalParameterOptimization}:
we apply the cross-sectional method to the template,
and use the estimated model parameters $\bldgr{\Omega}$
to initialize 
%$\{ \bldgr{\Omega}_t \}_{t=1}^T$.
%$\bldgr{\Omega}_t, \forall t$.
$\bldgr{\Omega}_t, t=1, \ldots, T$.
%%We set the 
%%using as
%with
%initial values of $\{ \bldgr{\Omega}_t \}_{t=1}^T$ 
%%to 
%%the parameter 
%%their
%set to
%the $\bldgr{\Omega}$
%values estimated 
%from the template 
%by the cross-sectional method,
%We segment the template with the cross-sectional model and set $\{ \bldgr{\Omega}_t \}_{t=1}^T$ to the estimated model parameters.
The proposed algorithm, which 
%iteratively updates 
interleaves updating
the latent variables $\bldgr{\theta}_0$ and $\fat{x}_0$ 
%and 
with updating
%subsequently 
the parameters
$\{ \bldgr{\Omega}_t \}_{t=1}^T$, is then run for five iterations, which we have found to be sufficient to reach convergence.

Based on initial pilot experiments on scans from the ADNI project\footnote{\url{http://adni.loni.usc.edu/}}
%(Cf. Sec.~\ref{sec:Experiments}), 
(distinct from the ones used in the experiments below),
we set the method's
hyperparameter values to $K_1=14K$ and $K_0=K$, where 
$K$ is the mesh stiffness in the existing cross-sectional method,
and 
% Find the best way to say how it is set.
$P_{0,k}$ to the number of voxels assigned to class $k$ in the
%cross-sectional
segmentation of the within-subject template.
% Looking at Andrew tuning this value 14 has almost same results as 10. Tuning was done by looking at hippo difference in ADNI dataset for 50 HC and 50 AD. So fig 2 results may be biased
%\todo{I'm starting to wonder if just using $10$ instead of $14$ would save us from this hairy biased results stuff, since that's just based on visual inspection. How annoying/fast would it be to re-run? Not first priority, though, but it is annoying}
%\answer{It takes days to compute everything unfortunately (145+87 subjects).}
%All the other hyperparameters of the model are set as in~\cite{Cerri2020}.
%\todo{Instead, so that the [cross-sectional?] stuff is implemented as in the cross-sectional paper.}

Our implementation builds upon the C++ and Python code 
%of~\cite{Cerri2020},
of~\cite{Puonti2016,Cerri2020},
and is publicly available from 
%***.
FreeSurfer\footnote{\url{http://freesurfer.net/}}. 
Segmenting one subject takes approximately 15
%$T$
minutes per time point
on an Intel 12-core i7-8700K processor with a GeForce GTX 1060 graphics card.

\section{Experiments and Results}
\label{sec:Experiments}

%\subsubsection{Datasets:}

In order
%to evaluate
to assess
whether introducing subject-specific latent variables 
%to the model 
leads to better longitudinal performance,
we compared the proposed method and the cross-sectional method on three different datasets:
%In order to validate our model and its cross-sectional counterpart, we used three different datasets:
%We 
%\todoDone{Here are at thee beginning of the section: Don't start with what you did, but why. Probably needs a separate introductory sentence}
%validated
%\todo{we didn't really ``validate'', which is hard to do since no ground truth, but rather ``compared''}
%the proposed model and its cross-sectional counterpart on three different datasets:
\begin{itemize}
    \item \textbf{Test-retest}~\cite{Biberacher2016}: This dataset consists 
    %\todoDone{here and elsewhere: ``It consists'' is weird to read.}
    of longitudinal T1-weighted (T1w) and FLuid Attenuation Inversion Recovery (FLAIR) scans of 2 MS subjects. For each subject 6 repeated scans were acquired from 3 different 3T scanners (Philips Achieva; Siemens Verio; GE Signa MR750) within 3 weeks. 
    \item \textbf{Achieva}: 
    %\todoDone{Think of a different name}
    This dataset consists of longitudinal T1w and FLAIR scans of 86 MS subjects. The subjects were scanned between 3 and 6 times (time between scans between 6 and 12 months) with a 3T Philips Achieva scanner at the 
    % We probably need to remove info for anonymization
    Department of Neurology, School of Medicine, at the Technical University of Munich in the context of the in-house cohort study on MS named TUM-MS.
    %***.
    All the subjects were diagnosed as relapsing-remitting MS. 
    %The subjects were divided into 3 age groups: $<$30 (n=25), 30-50 (n=52) and $>$50 (n=9).
    \item \textbf{ADNI}: This dataset consists of longitudinal T1w scans of 135 subjects randomly selected from the ADNI project.
    %\footnote{\url{http://adni.loni.usc.edu/}}
    Scanners from multiple sites were used to acquired the scans, and subjects were scanned between 2 and 6 times, with 6 or 12 months between scans.
    The subjects were divided into 3 groups: cognitively normal (CN, n=45), mild cognitive impairment (MCI, n=54), and Alzheimer disease (AD, n=36). 
\end{itemize}
We report results on 
the estimated volumes of
the following 26 
% main neuroanatomical 
regions: left and right cerebral white matter, cerebellum white matter, cerebral cortex, cerebellum cortex, lateral ventricle, hippocampus, thalamus, putamen, pallidum, caudate, amygdala and nucleus accumbens, as well as brain stem and lesions. To avoid cluttering, results for left and right structures are averaged. 

%\todo{``We registered all the T1w scans of each subject using an unbiased registration algorithm~\cite{Reuter2012}. We subsequently co-registered the FLAIR scans to the T1w. No other preprocessing step was performed on the scans.''}

%\subsubsection{Measures and metrics:}

%The experiments in this section concentrate on the following 26 main neuroanatomical regions: left and right cerebral white matter, cerebellum white matter, cerebral cortex, cerebellum cortex, lateral ventricle, hippocampus, thalamus, putamen, pallidum, caudate, amygdala, nucleus accumbens, brain stem and lesions. To avoid cluttering left and right structures are averaged. 

%We computed coefficients of variation (ratio of the standard deviation to the mean) for volumetric measurements of the same subject and of the same scanner for the Test-retest dataset.
%For computing temporal trajectories, for each structure and for each subject, we fitted a linear regression to the volumetric measurements at the various time points. We then computed for the Achieva dataset the ratio of the standard deviation of the residual to the intercept evaluated at the time of the first scan (called [does it have a name?] in the remainder of the paper).
%For the ADNI dataset, instead, we computed Annualized Percentage change (APC) as the ratio of the slope to the intercept evaluated at the time of the first scan. 

\subsubsection{%Regularization over time:
Temporal consistency:}
%\todoDone{Title no longer covers the content with the Achieva dataset included}

%In order to assess the ability of the proposed model to produce consistent segmentations across time,
%we compare its performance against the cross-sectional method on the Test-retest and Achieva dataset.

We wished to assess whether the proposed method is able to 
reduce non-biological variations in longitudinal 
volume
measurements,
both within the short 
($<3$ weeks) 
and longer 
($<6$ years) 
time intervals of the test-retest and the Achieva datasets, respectively.
For the test-retest dataset one can expect true biological 
%variations 
changes
to be minimal, and we therefore computed the coefficient of variation (ratio of the standard deviation to the mean) for each brain structure. The results, shown in Table~\ref{tab:3x3coefficients}, indicate that the longitudinal method indeed performs better 
in this respect
than the cross-sectional one for almost all the structures. 
%[, except for the cerebral cortex ].
%

For the 
%time span of the longitudinal scans of the 
Achieva dataset, 
%[ we assumed ] 
one may assume that the true 
%longitudinal volume change 
change in volume
of 
%each 
a 
structure
over the 
%time 
span of a few years
%resemble a linear trajectory, 
is approximately linear,
except for lesions whose temporal trajectory is affected by disease effects, with growing and shrinking lesions occurring at the same time.
%~\cite{Pongratz2019}.
%\unresolved{I can cite the paper, although it's more into lesion shrinking effects/prognosis than growing. I think Mark attached to show that lesion shrinking is still not understood and it's not a good idea to speculate on its trajectory.}
%\todo{ask Mark if OK like this}
We therefore 
%[ quantified how much volumetric measurements differed from a linear trajectory by fitting, ]
fitted, 
for each structure and for each subject, a linear regression model 
%[ to the estimated volumes ]
to the longitudinal volumes estimated by each method, 
%at each time point 
and 
%[ computing ]
computed
the ratio of the standard deviation of the residuals to the intersect (time of the first scan is taken as zero). The results are summarized in Table~\ref{tab:munich} and indicate 
%[, in accordance with the previous experiment,]
that the proposed model 
%has less variation in detecting linear trajectories
indeed
yields 
generally
%trajectories that are more linear
better results
in this respect.
%[ than the cross-sectional version ]
%[, except for the cerebral cortex]. 
%As expected, lesions have high variations for both models as they not 
%necessarily
%follow a linear trajectory. 
%\todo{to save space, could consider removing lesion result from text? (and even table?)}

\subsubsection{%\todoDone{sensibility?}
%Detecting group differences:}
Detecting disease effects:}

\begin{figure}[p]
%Image
%
%
%
%\begin{minipage}{0.75\linewidth}
%    \begin{center}
%    \includegraphics[width=\textwidth]{Images/segmentations/segExamplev2.png}    
%    \end{center}
%\end{minipage}
%\begin{minipage}{0.24\linewidth}
%    \caption{Example segmentation produced by the proposed method
%    %Segmentation
%     of a longitudinal scan with T1w and FLAIR contrast. 
%    %of the Achieva dataset
%    %with T1W and FLAIR as input.
%    \todo{Too small}
%    }
%    \label{fig:segmentationExample}
%\end{minipage}
%
%
%
\begin{minipage}{\linewidth}
\begin{center}
\begin{minipage}{0.6\linewidth}
    \begin{center}
        \includegraphics[width=\textwidth]{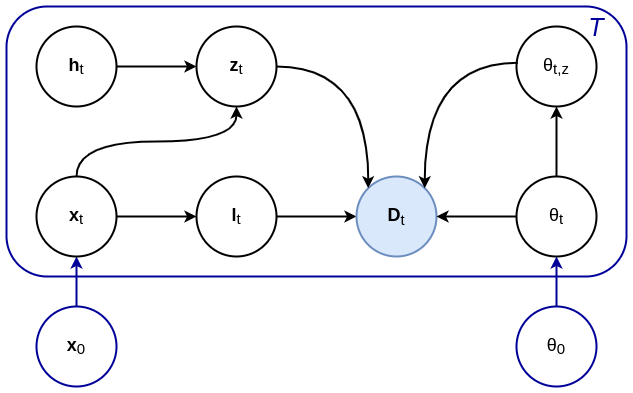}
    \end{center}
\end{minipage}
\begin{minipage}{0.38\linewidth}
    \caption{Graphical representation of the proposed model. In black the existing cross-sectional method of~\cite{Cerri2020} for each time point $t$; in blue the proposed additional latent variables for modeling temporal consistency between longitudinal scans with $T$ time points. %Shading indicates observed variables. The plate indicates repetition of the included variables.
    %\todo{two last sentences could potentially be removed for gaining space}
    %\todo{put caption beside figure to save space?}
    %\todoDone{$h$ should be $h_t$}
    %\todoDone{Plate and $T$ should also be in blue}
    }
    \label{fig:model}
\end{minipage}
\end{center}
\end{minipage}
%
%
% Exp 1 test-retest
%
%
\begin{minipage}{\linewidth}
\begin{center}
\begin{minipage}{.79\linewidth}
\begin{center}
\resizebox{\linewidth}{!}{%
\begin{tabular}{l||c|c|c|c|c|c||c|c|c|c|c|c||c|c}
 & \multicolumn{6}{c||}{Subject 1} & \multicolumn{6}{c||}{Subject 2} & \multicolumn{2}{c}{Avg} \\
 & \multicolumn{2}{c}{GE} & \multicolumn{2}{c}{Philips} & \multicolumn{2}{c||}{Siemens} & \multicolumn{2}{c}{GE} & \multicolumn{2}{c}{Philips} & \multicolumn{2}{c||}{Siemens}& \\
Structure & Cross & Long & Cross & Long & Cross & Long & Cross & Long & Cross & Long & Cross & Long & Cross & Long \\ \hline
Lesions & 3.79 & 3.90 & 4.78 & 3.79 & 6.27 & 3.39 & 2.84 & 4.04 & 2.57 & 1.95 & 2.35 & 1.22 & 3.77 & \textbf{3.05} \\
Cerebral white matter & 0.38 & 0.46 & 0.14 & 0.12 & 1.70 & 1.24 & 0.93 & 0.79 & 0.78 & 0.17 & 1.16 & 1.00 & 0.85 & \textbf{0.63} \\
Cerebellum white matter & 0.82 & 0.56 & 0.28 & 0.19 & 2.03 & 1.96 & 1.28 & 0.91 & 0.71 & 0.41 & 1.14 & 1.00 & 1.04 & \textbf{0.84} \\
Cerebral cortex & 0.53 & 0.50 & 0.50 & 0.44 & 1.50 & 1.84 & 0.60 & 1.08 & 0.30 & 0.40 & 0.74 & 1.10 & \textbf{0.70} & 0.89 \\
Cerebellum cortex & 0.38 & 0.34 & 0.31 & 0.36 & 1.13 & 1.25 & 0.61 & 0.63 & 0.42 & 0.36 & 0.86 & 0.57 & 0.62 & \textbf{0.58}  \\
Lateral ventricles & 1.10 & 1.03 & 1.37 & 0.38 & 1.11 & 0.57 & 3.73 & 2.95 & 1.62 & 1.10 & 2.71 & 1.24 & 1.94 & \textbf{1.21} \\
Hippocampus & 1.03 & 0.67 & 0.55 & 0.43 & 0.84 & 1.46 & 1.88 & 1.39 & 0.55 & 0.57 & 1.37 & 0.91 & 1.04 & \textbf{0.90} \\
Thalamus & 0.44 & 0.44 & 0.52 & 0.31 & 1.21 & 0.54 & 0.92 & 0.79 & 0.50 & 0.35 & 1.02 & 0.36 & 0.77 & \textbf{0.46} \\ 
Putamen & 0.58 & 0.14 & 0.84 & 0.76 & 0.70 & 0.71 & 0.51 & 0.31 & 1.10 & 0.52 & 1.44 & 1.08 & 0.86 & \textbf{0.59} \\
Pallidum & 2.25 & 1.35 & 2.04 & 1.58 & 3.94 & 1.31 & 2.77 & 2.39 & 1.12 & 0.77 & 4.35 & 1.84 & 2.74 & \textbf{1.54} \\
Caudate & 0.80 & 1.09 & 0.96 & 0.90 & 1.00 & 0.84 & 1.79 & 1.04 & 0.92 & 0.54 & 0.61 & 0.41 & 1.01 & \textbf{0.80} \\
Amygdala & 1.51 & 0.47 & 0.42 & 0.20 & 1.85 & 1.04 & 1.17 & 0.64 & 1.05 & 0.41 & 0.60 & 0.42 & 1.10 & \textbf{0.53} \\
Accumbens & 1.84 & 1.27 & 1.77 & 1.40 & 1.80 & 1.39 & 1.96 & 0.69 & 0.80 & 0.80 & 1.73 & 0.80 & 1.65 & \textbf{1.06} \\
Brain stem & 0.70 & 0.53 & 0.32 & 0.19 & 1.10 & 0.93 & 0.78 & 0.59 & 0.54 & 0.34 & 0.61 & 0.63 & 0.67 & \textbf{0.53} \\ \hline
Intracranial volume  & 0.30 & 0.06 & 0.13 & 0.18 & 0.53 & 0.56 & 0.46 & 0.06 & 0.22 & 0.08 & 0.52 & 0.36 & 0.36 & \textbf{0.22} \\
\end{tabular}
}
\end{center}
\end{minipage}
\end{center}
\begin{minipage}{\linewidth}
\captionof{table}{Coefficients of variation in [\%] 
on the test-retest dataset, 
%with input T1w and FLAIR, <- usually I would applaud this, but now (1) no space, and (2) not done for every figure
both for the proposed longitudinal (``Long'') and the cross-sectional (``Cross'') method.}
\label{tab:3x3coefficients}
\end{minipage}
\end{minipage}
%
%
% Exp 2 Achieva linear trajectories
%
%
\begin{minipage}{\linewidth}

\begin{minipage}{.6\linewidth}
\begin{center}
\resizebox{\linewidth}{!}{%
\begin{tabular}{l|c|c|l|c|c}
Structure & Cross & Long & Structure & Cross & Long \\ \hline
Lesions & 9.41 & \textbf{9.27} &
Cerebral white matter  & 0.56 & \textbf{0.31} \\
Cerebellum white matter  & 0.59 & \textbf{0.45} &
Cerebral cortex  & \textbf{0.54} & 0.59 \\
Cerebellum cortex  & 0.45 & \textbf{0.42} &
Lateral ventricles  & 2.04 & \textbf{1.85} \\
Hippocampus  & 0.69 & \textbf{0.55} &
Thalamus  & 0.51 & \textbf{0.49} \\
Putamen  & 0.70 & \textbf{0.42} &
Pallidum  & 1.23 & \textbf{0.78} \\
Caudate  & 1.01 & \textbf{0.80} &
Amygdala  & 0.70 & \textbf{0.40} \\
Accumbens  & 1.39 & \textbf{0.82} &
Brain stem  & 0.62 & \textbf{0.46} \\ \hline 
Intracranial volume  & 0.28 & \textbf{0.08} \\
\end{tabular}
}
\end{center}
\end{minipage}
\begin{minipage}{.39\linewidth}
\captionof{table}{
Average deviation from a linear trajectory in [\%] for 
%the 
%subjects' 
volumetric measurements in the Achieva dataset.
%Average of the variation 
%\todoDone{Unclear. I guess you mean deviation from linear trajectory? But perhaps it's easier to define a new term for std/offset in the main text so you have a word to use here?}
%\unresolved{Not sure how to call it since it doesn't have a ``real'' name. Is it not clear enough now?}
%of subject's 
%\todo{grammar? Article? Plural/singular genetive form?}
%volumetric measurements from a linear trajectory in [\%] for the Achieva dataset. 
%with T1w-FLAIR as input.
}
\label{tab:munich}
\end{minipage}
\end{minipage}

\begin{minipage}{\linewidth}
    \centering
    \begin{minipage}{0.75\linewidth}
    \begin{center}
        \includegraphics[width=\textwidth]{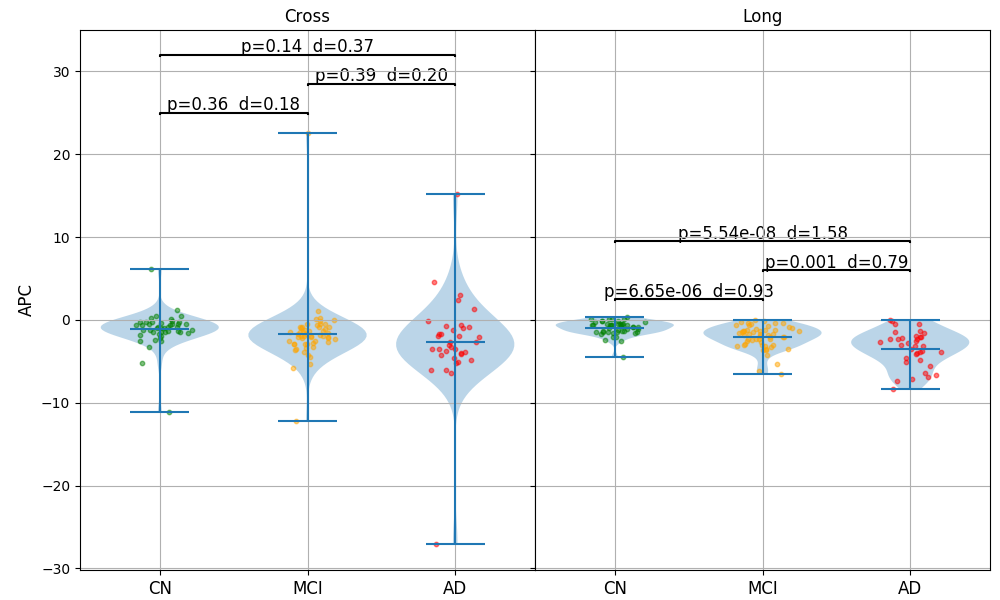}
    \end{center}
    \end{minipage}
    \begin{minipage}{\linewidth}
    \caption{Annualized percentage change (APC) in the volume of the hippocampus 
    %[ volumes of the hippocampus and the cerebral cortex ]
    for the three groups of the ADNI dataset.
    %both for the proposed longitudinal (``Long'') and the cross-sectional (``Cross'') method.
    Statistical significance was computed with a Welch's t-test and effect size with Cohen's d.
    % Maybe say that violins show means? <- No space for that -- not necessary for workhop paper.
    %\todoDone{Impossible to read the results in the figures without zooming 300\%. Stefano: I increased the font size. It's readable now with an a4 size view.}
    %\todo{Put caption beside figure to save space? Bigger font sizes here and in tables would be better}
    }
    \label{fig:AdniGroupAnalysis}
    \end{minipage}
\end{minipage}

\end{figure}

%%The test-retest reliability experiment
%The previous experiments inspected only 
%%one aspect of the longitudinal model, namely the 
%the proposed method's ability to produce consistent segmentations over time. We 
In order to 
%evaluate whether 
%demonstrate
ensure
that the proposed method is not simply over-regularizing,
%[ across time ], 
we
also assessed whether 
%the model can also 
it can 
%[better]
capture known group differences in the temporal evolution of certain brain structures
better than the cross-sectional method. 
%for different subject groups.
Towards this end,
we compared the annualized percentage change 
(the slope 
of a linear regression model
divided by its intersect)
%~\cite{Iglesias2016}
%\unresolved{Eugenio called this ``atrophy rates'', or ``slope in [\%] of baseline''. Annualized percentage change is how the Norwegian called this metric. I can't find a paper that exactly called it APC. Should I change the name?}
%\todo{never mind -- googling `` ``annualized percentage change'' brain volume'' shows it's a well-known thing in the literature. Let's keep that in mind for a journal version}
in the
volume of the hippocampus 
%[ volumes of the hippocampus and the cerebral cortex
%\todo{Since you don't say anything about hippo in tuning part, can save space by skipping cortex?}
%] 
between the CN, MCI and AD groups in the ADNI dataset.
%
%hippocampus and cerebral cortex. % I need to find a paper saying atrophy in AD for hippo and cortex <-- No you don't -- that's very well established now.
%We computed Annualized Percentage change (APC) as the ratio of the slope to the intercept evaluated at the time of the first scan, with slope and intercept computed by fitting a linear regression to the subject's volumetric measurements. 
%Fig.~\ref{fig:AdniGroupAnalysis} shows APC values for the three different groups of the ADNI dataset. 
The results, 
shown in Fig.~\ref{fig:AdniGroupAnalysis},
indicate that the longitudinal method 
can
indeed 
detect group differences 
better this way.
%[ than the cross-sectional version ]
%[ that the cross-sectional method could not find, or detect differences with a higher statistical significance.
%\todo{only needed if we keep cortex results}
%] 

\section{Discussion and conclusion}
\label{sec:Discussion}

% Structure of the discussion
% 1) What we proposed here
% Longitudinal model 
% contrast adaptive and no requirement on the number of time points
% subject specific atlases and intensity latent variables
%
% 2) Validation
% longitudinal model better than cross sectional model in:
% 1) scan rescan volume differences (lower differences)
% 2) group comparisons
%
% 3) Limitations
% no ground truth for Achieva dataset
% biased results for Adni hippocampus
%
% 4)Future work
% Validate on more MS data (Munich) / with MS groups or EDSS progression
% Better tuning of the hyperparameters

In this paper we have proposed a novel method for the segmentation of longitudinal brain MRI scans of patients suffering from MS. 
The method is based on an existing cross-sectional method for simultaneous whole-brain and lesion segmentation, 
and leverages subject-specific latent variables to encourage segmentations across time points to be similar to each other. 
%It is very generally applicable, as it does not make any prior assumptions on the scanner, the MRI protocol, or the number and timing of longitudinal follow-up scans.
%
Preliminary results indicate that 
%the proposed method 
it
is able to produce more consistent and reliable segmentations compared to the cross-sectional version, while 
% being able to better detect group differences. 
being more sensitive to group differences.
%
%has better reliability and sensibility compared to its cross-sectional counterpart. 
% I guess here something about the possible biased results of the adni dataset for hippocampus
%The results presented here are limited by 
%%the absence of 
%%%ground truth 
%%group differences
%%for the Achieva dataset, 
%in the employed longitudinal MS dataset,
%restricting the analysis of detecting disease effects on non-MS patients. % (ADNI dataset).
Future work will involve an extensive analysis of disease progression in different MS 
patient groups, as well as a more careful tuning of the hyperparameters of the model.
%\todo{Stefano: since I've corrected this before, I'll just point it out this time: "patient groups" instead of "group patients"}

\subsubsection{Acknowledgments:} %***
\label{sec:Acknowledgments}

% This need to be anonymized
%\begin{comment}

This project has received funding from the European Union's Horizon 2020 research and innovation program under the Marie Sklodowska-Curie project TRABIT agreement No 765148,
%Mark M{\"u}hlau was supported by
the German Research Foundation
%(Priority Program SPP2177, Radiomics: Next Generation of Biomedical Imaging) -- 
No 428223038, and 
%the National Institute Of Neurological Disorders and Stroke under project number
the NIH NINDS No
R01NS112161.

%\end{comment}

\bibliography{refs}

\end{document}